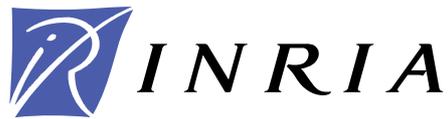

# On Frequency Optimisation for Power Saving in WSNs: Finding Optimum Hardware Timers Frequencies

Andreea-Maria Picu — Antoine Fraboulet — Eric Fleury

# On Frequency Optimisation for Power Saving in WSNs: Finding Optimum Hardware Timers Frequencies

Andreea-Maria Picu, Antoine Fraboulet, Eric Fleury

**Abstract:** Wireless Sensor Networks research and demand are now in full expansion, since people came to understand these are the key to a large number of issues in industry, commerce, home automation, healthcare, agriculture and environment, monitoring, public safety etc. One of the most challenging problems in sensor networks research is power awareness and power-saving techniques. In this research report, we have studied frequency scaling as a power-saving technique. In particular, we analysed the close relationship between clock frequencies in a microcontroller and several types of constraints imposed on these frequencies, e.g. by other components of the microcontroller, by protocol specifications, by external factors etc. Among these constraints, we were especially interested in the ones imposed by the timer service and by the serial ports' transmission rates. Our efforts resulted in a microcontroller configuration management tool which aims at assisting application programmers in choosing microcontroller configurations, in function of the timing needs and constraints of their application.

**Key-words:** hardware node configuration, power awareness, sensor network node, wireless sensor networks, energy saving, frequency scaling

# De l'optimisation des fréquences pour économiser de l'énergie dans les capteurs sans fil: Calcul des fréquences optimales pour les temporisateurs matériels

**Résumé :** La recherche dans le domaine des réseaux de capteurs sans fil et la demande pour ce type de réseaux sont en pleine expansion, depuis que les gens ont compris que ceux-ci représentent la clé d'un grand nombre de problèmes dans les domaines de l'industrie, du commerce, de la domotique, de la santé, de l'agriculture et de l'environnement, de la surveillance, de la sûreté publique etc. L'une des problématiques les plus ambitieuses dans la recherche sur les réseaux de capteurs est la gestion de l'énergie et les techniques d'économie d'énergie. Dans ce rapport, nous avons étudié une technique d'économie d'énergie en particulier, à savoir la réduction judicieuse de la fréquence. Nous avons notamment analysé le lien très étroit entre les fréquences des horloges dans un microcontrôleur et les nombreuses contraintes imposées à ces fréquences, par exemple par d'autres composants du microcontrôleur, par des spécifications d'un protocole, par des facteurs externes etc. Parmi ces contraintes, nous nous sommes particulièrement intéressés à celles imposées par le service de temporisation et par les débits de transmission des ports série. Nos efforts ont eu pour résultat un outil de gestion des configurations, dont le but est d'aider les développeurs d'applications dans leur choix des configurations du microcontrôleur, en fonction des besoins et des contraintes spécifiques de leur application en termes de temporisation.

# Acknowledgements

I would like to thank both my master's thesis directors for trusting my intellectual capacities and for giving me the opportunity to put the knowledge gathered through the years of study into practice within their research team. From the beginning I have experienced a great hospitality for which I am most grateful, and which has made this experience a real pleasure.

Special acknowledgements go to Mr. Antoine Fraboulet, who has always been very available and understanding and who offered valuable ideas and guiding directions to overcome all the issues encountered during this project.

It is not easy to name and thank every person who has been, in one way or another, involved in this internship, without running the risk of forgetting somebody. I think it is therefore wise to send out a general thanks to those persons who have made their contribution to this project and without whom, I would not have been able to accomplish my goals.

# Contents

# List of Acronyms

| | |
|---|---|
| A/D | Analog-to-Digital, |
| ALU | Arithmetic Logic Unit, |
| AODV | Ad-hoc On-demand Distance Vector, |
| API | Application Programming Interface, |
| APS | Application Support Sub-layer, |
| | |
| CAP | Contention Access Period, |
| CFP | Contention Free Period, |
| CPU | Central Processing Unit, |
| CSMA-CA | Carrier Sense Multiple Access - Collision Avoidance, |
| | |
| DCO | Digitally Controlled Oscillator, |
| DFS | Dynamic Frequency Scaling, |
| | |
| FSM | Finite State Machine, |
| | |
| GCD | Greatest Common Divisor, |
| | |
| IEEE | Institute of Electrical and Electronics Engineers, |
| ISA | Instruction Set Architecture, |
| | |
| LCM | Least Common Multiple, |
| LIFS | Long Interframe Space, |
| LPM | Low Power Mode, |
| | |
| MAC | Medium Access Control, |
| MCU | Microcontroller Unit, |
| MINLP | Mixed-Integer Nonlinear Programming, |
| | |
| NLP | Nonlinear Programming, |
| OS | Operating System, |
| OSI | Open Systems Interconnection, |
| | |
| RAM | Random Access Memory, |
| RF | Radio Frequency, |
| ROM | Read-Only Memory, |
| RREP | Route Reply, |
| RREQ | Route Request, |
| | |
| SHR | Synchronisation Header, |
| SIFS | Short Interframe Space, |
| | |
| WSN | Wireless Sensor Network, |
| | |
| ZDO | ZigBee Device Object, |

# 1 Introduction

This report investigates the use of advanced power saving methods as part of the energy management scheme of a specific kind of mobile battery-powered computing devices, also known as sensing devices. It is an attempt to make use of a greater part of the realm of possibilities offered by hardware, aiming at the replacement of the single system-wide power setting with a per application setting. To conduct the analysis, we use the TI MSP430 microcontroller, but the approach we took allows the reproduction of the study for other microcontrollers.

## 1.1 Motivation

In recent years, a lot of research has been conducted on an important class of collaborating objects, i.e., wireless sensors, which will constitute the infrastructure for the ambient intelligence vision. This new kind of embedded systems has great potential for many applications, for example surveillance, disaster relief applications, environment monitoring, emergency medical response and home automation etc.

Given its numerous domains of application, industry is also beginning to express its interest in wireless sensors. However, the great mismatch between research at different levels (application, network, node) has forced industry to be reluctant to use research results. This study is part of greater initiative [26] to narrow this mismatch.

Power management in WSNs is a very rich area, since the matter is of utmost importance in this field. Motivations for the extensive research carried out in this field are numerous. Firstly, the battery is generally not replaceable, due to the randomness of the sensing device's position and sometimes also to the dangerousness of the sensing field. Therefore, battery lifetime is synonym of sensor lifetime and must be extended as much as possible. Secondly, the progress made in battery capacity, lifetime and size is at best limited as compared to the one in processing power, storage capacities and size. Lastly, with performance constraints ever-increasing, power management is guaranteed to always need improvement.

Techniques for the administration of energy resources and for the optimisation of power consumption have been proposed at almost all levels of a system from the ISA level to the operating system level [11]. In the field of WSNs, schemes for power saving often lie in communication protocols, since one of the most energy-expensive operations is data transmission.

However, computation, memory access, or peripherals are also power greedy. Computation alone consumes almost as much energy as communication, especially in leader nodes or sinks. Indeed, the role of leader nodes is to collect and relay information, as opposed to sensing, which makes the benefits of power aware communication protocols much less significant than power aware computation for this type of nodes.

All this is to say that, in order to account for application specificities, a holistic approach to the optimisation of power consumption is essential. Power saving schemes incorporated in the OSI model are reductive, whereas a direct, fine-grain control over underlying hardware would allow the application to dictate when power can be saved.

The following questions must be answered. Is it feasible for an application to directly control the energy consumption of the hardware platform at runtime? How can user and/or application information be incorporated into the power management scheme? Is it possible to derive an optimal hardware configuration with this information? These questions are addressed in this report.

## 1.2 Objectives

The goal of this study is the exploration of different approaches to application-specific power administration, such as application-driven frequency scaling in combination with enhanced hardware resources allocation. Different applications have different needs and are influenced in different ways by energy management policies. It may even happen that the same application have different requirements throughout its lifetime. It is thus an important matter to investigate the way in which the application itself can influence or even control power management.

Figure 1 presents our approach to WSN energy management, addressing this matter explicitly. The approach consists of several building blocks, all of which will not be addressed in this report. The first building block comprises software requirements and needs. In order to enable the application to influence power management, the application's requirements must be derived and this is the role of the application framework. The protocol stack and its needs also have an important part in our scheme.

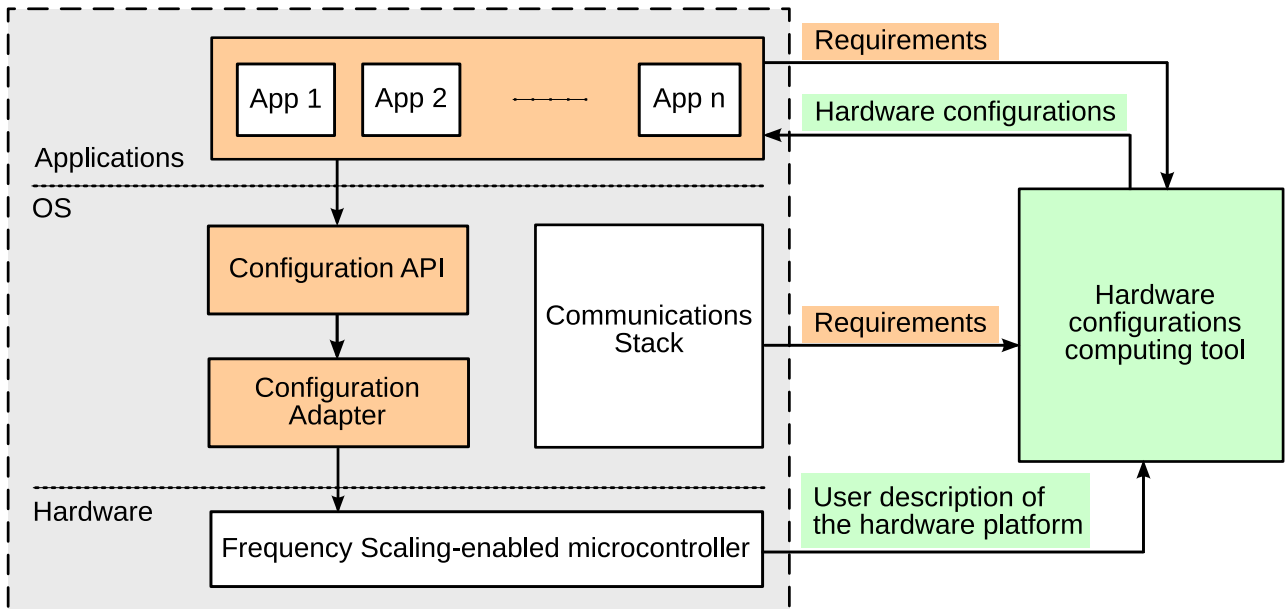

Figure 1: Outline of our power management approach

The description of the hardware platform combined with the above mentioned requirements and needs will allow the computing tool to calculate a certain number of power-efficient configurations which will perfectly match requirements to available resources. These hardware configurations are then sent back to the application framework. The latter will hence be able to switch among hardware configurations in function of the applications' computational activity, of external events etc. The switching is handled by the configuration API and adaptor.

The focus of this report is the description of hardware resources and the hardware configurations computing tool (green colour in Figure 1). The rest of the approach (red colour) is left as future work and briefly discussed in Section 6.

Our study concentrates on resource reallocation, since it is the key to allowing the application a fine-grain control over hardware resources without placing too much responsibility on the programmer's shoulders. In contrast to other studies on WSN energy management, minimising the power consumption is not the only and primary goal, as the report considers its relation to the improvement of hardware resources usage.

## 1.3   Outline

The remaining of this document is divided into four chapters, as follows. Section 2 introduces the background needed and some related issues. Section 3 presents the various hardware description methods used throughout our study and explains the usage of these descriptions. The new resource allocation algorithm that we developed and the configurations computation tool are depicted in Section 4. Finally, Section 5 presents a case study (the ZigBee protocol stack) and evaluates the improvements of our scheme as compared to common techniques. Perspectives and future work will be discussed in conclusion.

# 2 Background and Related Work

The first obvious prerequisite to understanding WSN power management is to have at least a basic understanding of WSNs and their challenges. Most importantly, current power saving mechanisms and their implications have to be understood. Therefore, this chapter is dedicated to laying the groundwork for the rest of the study.

First, a brief presentation of WSNs will be made. Then, we will overview some power saving methods and the most famous WSN operating systems. And finally, the last section will relate our work to an important aspect of power management: battery models and discharge.

## 2.1 Brief Presentation of WSNs

A sensor is a device which measures a physical quantity and transforms it into meaningful information. It is a basic element of data acquisition systems, which are needed for observation and control in numerous scientific and research applications.

In the past, the only solution for transporting the data from the sensor to the central controller was wiring, which had important drawbacks, like its price and its overall dimensions. Nevertheless, thanks to the huge progress in wireless communication this wiring can now be eliminated to form Wireless Sensor Networks (WSNs).

WSNs are a very popular theme at the moment, both for the research world and for industry and are considered as one of the ten new technologies which will deeply change the world and people's lives. Technological progress has allowed the manufacture at a reasonable cost of tiny devices consisting of: a sensing unit, a processing and storage unit, a wireless transceiver unit and a power unit. Therefore, these devices are veritable embedded systems. The rollout of several such devices, in order to autonomously collect and transmit data to one or several collection points forms a wireless sensor network.

WSNs are a special type of ad hoc network (or infrastructure free network). The nodes of this type of network are mostly sensors whose positions are not necessarily predetermined. They are randomly scattered through a geographical zone, also called a sensing field, which defines the terrain of interest for the sensed phenomenon.

The sensed data are transported by way of multi-hop routing to a collecting point, also called a sink. The latter may be connected to the user through the Internet or through a satellite link. Thus, the user can make requests to other nodes on the network, specifying the type of data required, and collect the sensed data through leader nodes.

WSNs bring a new and very interesting perspective: that of networks capable of autoconfiguration and of self management, eliminating the need for human intervention. Moreover, performance criteria for a sensor network are different from classical networks, therefore, new solutions are needed. In fact, wireless sensors are intended to become ordinary objects which are easy to use. The network must be transparent to the user.

The realisation of this type of network requires the use of techniques that were originally developed for ad hoc networks. However, most of the protocols developed for the latter are not transposable as such to sensor networks, since the issues are different. From a technical point of view, sensor networks raise new challenges like routing, power management, autoconfiguration, dissemination and data collection, rollout etc.

## 2.2 Current Power Saving Mechanisms

Since this section will look into embedded systems and hardware platforms, a brief introduction to microcontrollers is necessary [29, 3]. A microcontroller is a computer-on-chip, meaning that in addition to the components of any microprocessor chip (ALU, control unit, registers), it contains additional elements such as RAM and ROM. In short, a microcontroller is a single integrated circuit, commonly with the following features: CPU (4-bit to 64-bit), serial ports, RAM, ROM, A/D converters, peripherals (e.g., timers, watchdog) and one or more clock generators.

Figure 2 presents an example of architecture, with a zoom on the clock system. There are generally several clocks in a microcontroller and, except for the CPU (which always uses the master clock), all other elements must choose among a set of clocks. These clocks themselves may be the result of a multiplexing of several clock generators, as shown in Figure 2. Reducing the power consumption by scaling the frequency of the clock generators is therefore a very hard task, since it will most likely influence the entire clock system.

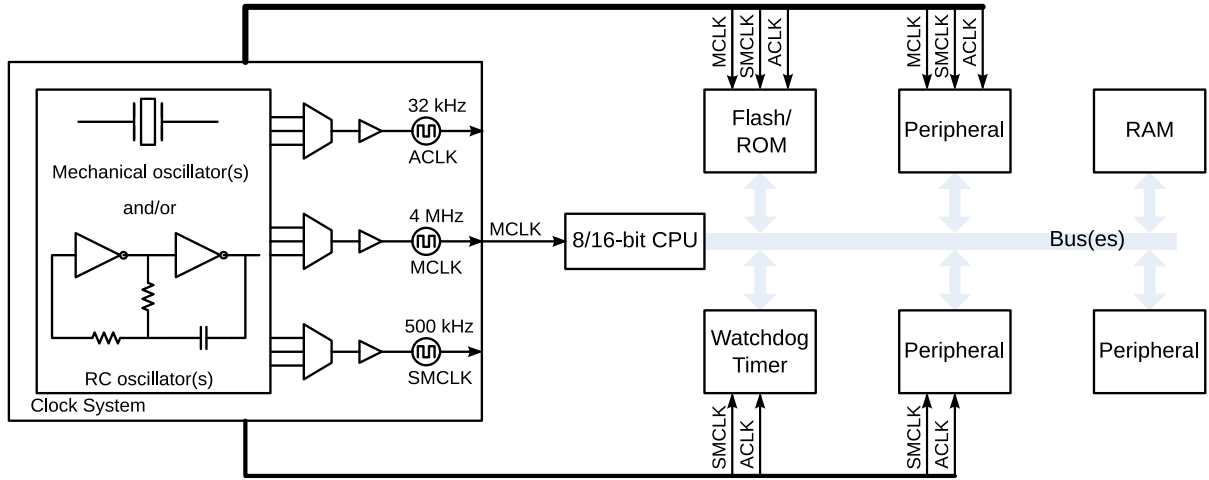

Figure 2: Example of microcontroller architecture

Some peripherals, for example timers, are not easy to manage even when the clock frequency is not modified. This is why embedded systems often use only one hardware timer to implement a list of software timers, even though up to sixteen hardware timers are available. In a nutshell, scaling the frequency of clock generators in a microcontroller requires the setting up of a mechanism to control all parts affected by the scaling.

To the best of our knowledge, there has been little work on controlling the power consumption of an entire hardware platform for a sensor node [10]. Yet, this is very important for embedded systems, like WSN nodes. Previous studies on this subject fall into two categories: the ones that consider the whole platform, but do not offer enough control (e.g., low power modes (LPMs)) and the ones that only investigate certain parts of the platform (e.g., energy-aware communication protocols, CPU frequency scaling).

LPMs [12] vary the degree to which the microcontroller unit (MCU) is aware of its surroundings and the clocks that the MCU keeps running. The MCU lowers power consumption partly by shutting off external and internal oscillators. Unused peripheral modules on the MCU are also de-activated to save power. Although LPMs are very effective, their design follows a on/off rationale. For applications which have very small execution times (1-2 seconds), the time and energy overhead for putting the system to sleep and waking it up may exceed the expected benefits. Thus, for this type of applications other methods are more adequate, e.g. frequency scaling and voltage scaling. Voltage scaling is not an available feature of most microcontrollers used in sensing devices.

Thus, our attention turned to frequency scaling. Frequency scaling is a technique where the processor clock is reduced by some multiple of the maximum, permitting the processor to minimise the energy dissipation linearly, as shown in equation 1 ($P$ is the power consumed, $\alpha$ is the activity factor, $C$ is the switched capacitance, $V$ is the supply voltage, $F$ is the clock frequency). It allows to save energy even when it is not profitable to go into LPM at the expense of reduced performance. Scaling down the frequency will reduce the microcontroller's power dissipation in Watt but cannot save the energy in Joule (see equation 2) consumed by a task, because for a given task, $Ft$ is a constant (the system will take a longer time to execute its workload at a lower frequency).

But if the system is under-utilised this method, alone or in combination with LPMs, will save a significant amount of power and energy. The authors of [11] find that "although dynamic voltage scaling renders the lowest energy dissipation for most microcontrollers, it is not always dramatically better than using a combination of dynamic frequency scaling and the inbuilt power down modes, which is much less expensive to implement". Moreover, reducing the power dissipation will have a significant positive impact on battery capacity, as will be discussed in Subsection 2.4. Frequency scaling is also essential if we plan to use voltage scaling in the future. This is very likely to happen, since microcontrollers make rapid progress and the feature will surely be available in the near future for chips used in WSNs.

$$P = \alpha C V^2 F \qquad (1)$$
$$E = Pt \qquad (2)$$
$$= \alpha C V^2 Ft$$

A lot of research has been conducted on dynamic frequency scaling (DFS), since the time of its introduction around 1994, generally for high-end computing platforms (32-bit and 64-bit processors) [17]. However, there is one study of this technique for microcontrollers [11], proving that these advanced power saving methods are applicable to such devices. However, all these studies suffer from the drawback of only analysing frequency scaling as applied to the core processor or microcontroller and only at the circuit level or at most at the operating system level.

In addition to these components, in a typical embedded system, like the ones based on the TI MSP430 [12], the processor is also attached to various peripherals e.g., timers, serial ports, flash memory etc. Apparently, frequency scaling also has an affect on the energy consumed by these peripherals. On one hand, decreasing the operating frequency of the processor increases the time that these peripherals are turned on, thereby resulting in an increased energy consumption, which overshadows the energy savings gained from the processor consumption. On the other hand, decreasing the operating frequency of the entire platform, including peripherals, will eliminate excesses and save energy. Thus, for an embedded system like the TI MSP430 platform, it is essential to consider the application of frequency scaling to the peripherals as well.

In our study we analyse this method as applied to the entire hardware platform (including sub-systems on the chip and peripherals) and more importantly, its consequences on the WSN applications and the operating system. It is the study of these consequences that led us to believe that a better timer service management will greatly contribute to the optimisation of the power consumption of the entire platform.

## 2.3 Timer Management in WSN OSs

As pointed out above, our holistic approach to frequency scaling is closely related to the timer service and to timer management. This is why a survey of timer management in popular WSN operating systems, like TinyOS, Contiki and SOS, was necessary [29, 1]. An overview of these follows below.

A critical part of a WSN operating system is offering a reliable, powerful, and efficient timer service. This service must provide a standard interface to an arbitrary number of timers, in order to support portable, composable high-level services. Timer rates vary from a few events per day to sampling rates of 10 kHz or even higher. Finally, the timer service must allow the sensing device to be placed in a LPM between timer events.

Microcontrollers used in sensor platforms come with a wide variety of hardware timers. For instance, the Atmel ATmega128 has two 8-bit timers and two expanded 16-bit timers (each with several compare units), while the TI MSP430 (according to models) comes with two cascadable 8-bit timers and two 16-bit timers, summing up to 12 compare units. Both microcontrollers also carry watchdog timers.

All of these timers come with different clocking options, compare and external event capture registers etc. But most of the time only a minimal percentage of this plethora of options and features is used, as we will see below. The general trend is to fix the main clock frequency and to have all software timers be served by one hardware timer. This is known under the name of timer virtualisation and it is not very energy efficient.

### 2.3.1 TinyOS 2.x

TinyOS 2.x developers believe that a standard interface cannot provide a consistent view of the hardware diversity. Therefore, they chose to use the telescoping abstraction principle for TinyOS 2.x's timer subsystem. At the top-level are virtualised and shared timers with a standard, limited interface (the `Timer` interface). These virtualised timers are statically allocated to different services. Underneath, there are microcontroller-specific interfaces to the hardware timers. These timers are physical and dedicated, e.g., providing the virtual timers, or doing cycle-counting for benchmarking purposes [14, 24].

The granularity of the timer depends on the component providing the `Timer` interface[1]. For now, only the millisecond timer has been implemented, but two other types have been defined (the $\mu$s and the 32 kHz timers) and may be provided by platforms. Timer virtualisation is ensured by a module using the parametrised interface concept. The concept is similar to a hash function: each timer has a unique identifier which allows the system to retrieve it in constant time.

In the case of the TI MSP430, a module exists that exposes almost entirely the functionalities provided by the TI MSP430 timers A and B, including clock source selection and division. But, these extraordinary functionalities are only used at setup time, when the master clock frequency is set to a fixed value (4 MHz), which will never be changed. Every time a timer expires, the timer list (or "hash table") is skimmed through and the next timer is set in one of timer A's or timer B's compare units.

---

[1] See Appendix A for a diagram of the components.

To sum up, although exposing most of the hardware functionalities is a huge progress, TinyOS 2.x still uses a unique fixed frequency and only one of the dozen available hardware timers.

### 2.3.2 Contiki

Contiki offers four types of timers: `timer` (passive timer, follows only its expiration time), `etimer` (active timer, sends an event when it expires), `ctimer` (active timer, calls a function on expiration), and `rtimer` (real-time timer, calls a function at an exact time; not implemented for the TI MSP430 yet). In reality, the first two are the main types, since `ctimer` and `rtimer` actually make use of an `etimer` to achieve their missions [7, 6].

Timer A from the TI MSP430 is used as clock. Its source is the sub-system master SMCLK and it cannot be changed. There is no reusable frequency setting function, everything is done in the setup function. Event timers are organised in a linked list which is skimmed through at each update. Regular timers are used by verifying their state (expired or not), which is calculated on demand by querying the current time and subtracting the timer interval from it.

### 2.3.3 SOS

In SOS, the developers have used one hardware timer and allow an unlimited number of timer requests up to memory limitation. The timer service uses the hardware timer to account for the number of ticks since timer request. The resolution of timer tick is $\frac{1}{1024}$ seconds. To schedule a one second timer, an application will need to request 1024 ticks [18].

SOS developers have chosen to structure timer requests according to the delta value of the previous request. Instead of maintaining the timeout value in absolute value, the timeout value is represented as the value relative to previous timeout. For example, if module X requests a timeout of 1000 ticks and module Y requests a timeout of 1500 ticks, module X's request will be stored as 1000 and module Y's request as 500. This way, only module X's request needs to be decremented on updates. This makes the implementation a little bit complicated since each new timer must be inserted at the right place in the delta values linked list, but it is more resource efficient.

There are a number of timers that are pre-initialised in the system and timers are organised in several double-linked lists (pre-allocated timers, initialised timers, periodic pool) to allow for memory re-usage instead of freeing and malloc'ing all the time. The delta queue is updated on every hardware interrupt from TI MSP430's timer B. There is no reusable frequency setting mechanism, everything is done in the setup function (currently 4 MHz).

In a nutshell, the analysis of these OSs proved that at present, timer management systems are rather rudimentary. They all use only one hardware timer to provide a certain number of software timers. Although it is neither resource efficient, nor energy-aware, this is a convenient and easy way to handle timers. It assumes that the clock frequency never changes and this is the reason why frequency scaling cannot be used in combination with this timer management method. However, microcontrollers come with a multitude of hardware timers that can also be used in the software timer system. The optimisation of the use of hardware resources, like timers, would result in lower operating frequencies for the microcontroller and would thus save power and increase battery capacity.

## 2.4 Battery Discharge Behaviour

This report would not be complete without a run through the results of the latest studies in the field of battery discharge behaviours. Indeed, the energy drawn from a battery is not always equivalent to the energy consumed in device circuits. Therefore, battery discharge is an important point, since it can give an a priori indication of the effectiveness of a power-saving scheme.

The result which is most important to our study, [16], proves that peak power drawn from a battery rather than average power determines that battery's capacity. Consequently, reducing the peak power of a mobile system will increase the battery life by more than reducing the idle power, even if both reductions result in the same average power. An important method of reducing peak power is trading power for performance, particularly scaling the clock frequency down.

Another study that is relevant to our work is [21]. The authors explain the rate-dependency of a battery's capacity. They state that battery capacity decreases as the discharge rate increases and illustrate the phenomenon by a simplified example. In short, for the energy-generating chemical reaction to take place in the battery, active species need to be present at the electrode surface. The active species are replenished by diffusion from the bulk of the electrolyte. However, the higher the load current, the harder it is for the diffusion process to keep up with the reaction process. So, when the concentration of active species at the electrode surface drops below a certain threshold the battery no longer produces energy, even though charge is still available. The charge will eventually re-become available after a sufficiently long time. However, at sufficiently low discharge rates, the battery will behave like an ideal energy source and the phenomenon depicted above can be avoided.

To conclude this section, it is important to keep in mind the results of the above research on battery discharge behaviour when attempting to design new power saving techniques. In our case, these studies reinforced our belief that minimising the power drawn from a sensing device's battery will bring significant benefits.

## 2.5 Summary

This chapter briefly presents wireless sensor networks and current power saving techniques applied in this field. We concluded that frequency scaling would be a novel approach to power management in WSNs. Battery discharge behaviour and energy and power equations lead to believe that the above approach will have a significant effect on power consumption. Frequency scaling is closely related to resource allocation, in particular to timers, therefore we also studied timer management in popular WSN operating systems.

# 3 Hardware Descriptions

For a better allocation of the hardware resources, the allocation scheme must have knowledge of the available resources. Also, a holistic frequency scaling approach requires detailed knowledge of all the dependencies between clocks and microcontroller sub-systems and also among sub-systems themselves. This is why we decided to describe the available hardware resources in a way that will be easily manipulated by a software program.

As our study advanced, we used several ways of representing available hardware resources. Below are presented these different types of hardware descriptions and the reasons we considered them. But first, a common point of all our methods of describing hardware is introduced.

Although the TI MSP430 was chosen for this study, the approaches we took do not make any assumptions on the microcontroller or on the operating system, therefore generality is preserved. For the purpose of our analysis, we split the microcontroller into several blocks, corresponding to sub-systems sharing the same clock. Figure 3 shows these blocks and the clocks that each block can choose from, as well as interactions between blocks.

This hierarchical organisation is very convenient, no matter the way we choose to describe our hardware, since it separates the parts of the microcontroller using different clock sources. In the context of frequency scaling, this is an essential property of the description.

## 3.1 Hardware Dependency Graph

The first description took the form of a dependency graph, with registers being vertices and clocks being edges. Given the above property, the dependency graph is a hierarchical graph: each block corresponds to a sub-graph. An example of sub-graph is shown in Appendix B. In this diagram, each shape corresponds to a different type of register. The possible types are:

- on/off switching register – double circle,
- selection register – inverted trapezium,
- decision register – diamond,
- division register – inverted triangle,
- clock source – hexagon,
- other – box.

The diagram in Appendix B is the automatically generated `.dot` representation of the graph data structure containing that information. The process to obtain the hardware resources information under that form is described below.

The relevant microcontroller information is gathered in the form of an XML file, written by the user. Currently, we completed the XML hardware description for the TI MSP430. When writing this file, a specific grammar must be respected (see Appendix C). The resulting XML file will be parsed, using the Xerces C++ SAX parser [8], and converted into a graph data structure that can be easily manipulated by software, using the Boost Graph Library [5]. As pointed out before, this is a general approach and any other microcontroller can be described using this grammar and parsed to obtain the graph structure.

This description was intended to support a slightly different approach than the one introduced in the beginning of this report. The idea was to allow the application to dynamically control power management at runtime through an Application Programming Interface (API). The initial goal of this representation of the hardware resources was to help in the propagation of clock frequency changes throughout the microcontroller in function of the sub-systems which depend on the modified frequency. The graph would be used to generate a state space (with a state being composed of the values of all the configuration registers of the microcontroller at a given time) containing all the valid states of the microcontroller.

A program implementing a Finite State Machine (FSM) would perform the transitions between states and reconfigure the hardware. Hardware reconfiguration would thus be transparent for the application, which would simply use the interface to the above software program. However, we soon realised that this solution is too complex (the size of the state space would be enormous) and thus not feasible. Moreover, the amount of energy consumed for state transitions would exceed the one saved by this operation, due to the large amounts of computation and memory involved.

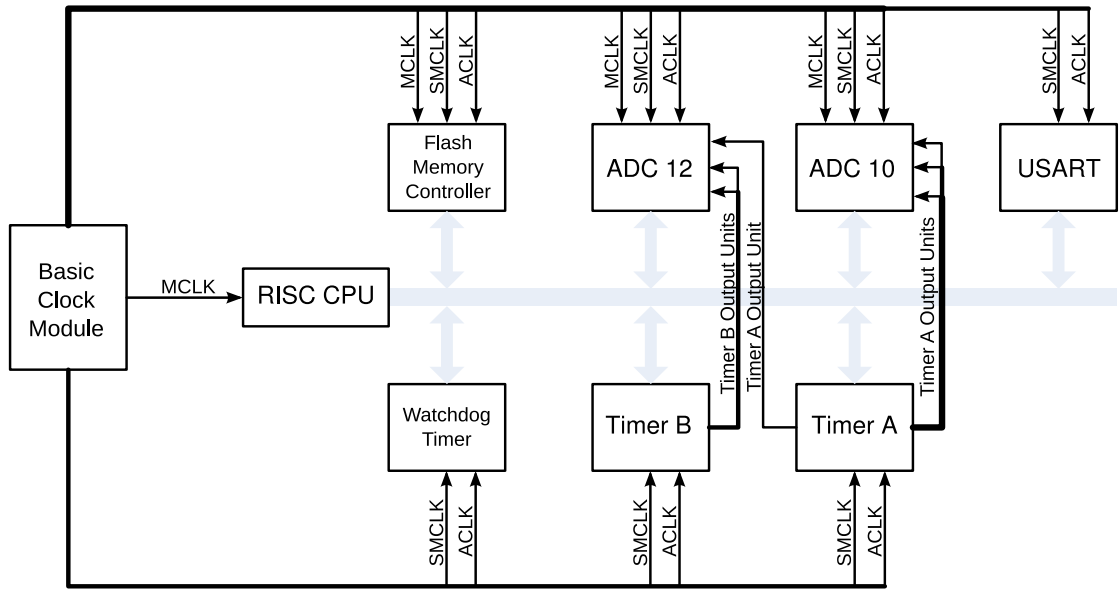

Figure 3: Sub-systems of the microcontroller sharing the same clock

## 3.2 Nonlinear System of Equations

The second description is analytic. We chose this particular kind of modeling, hoping that it will eventually allow us to find a minimum frequency (given a set of constraints) and the associated hardware configuration, by simply solving a system of equations. We kept the hierarchical separation mentioned above, for clarity. Similarly, to the first description, we categorised hardware configuration registers according to their types. To describe each type of register we used either an equation or a system of equations.

User or application imposed constraints (more about this in Section 5) allow us to assign values to certain parameters and thus solve the overall system of equations for our main unknowns: clock frequencies. The goal is to compute the values of the hardware registers which minimise these frequencies.

Similarly to the previous description, we classified registers into several types. Two simple examples of these types and their associated equations follow[2]:

- selection register or selector:
$$\begin{cases} a_1 X_{in} + a_2 Y_{in} + a_3 Z_{in} = T_{out} \\ a_1 + a_2 + a_3 = 1 \\ a_1, a_2, a_3 \in \{0, 1\} \end{cases}$$

  where $X_{in}$, $Y_{in}$ and $Z_{in}$ are input frequencies, $T_{out}$ is the output frequency and $a_1$, $a_2$, $a_3$ are parameters. Frequencies are also bounded either by the oscillating frequency of a crystal or by the range of a Digitally-Controlled Oscillator (DCO).

- division register or divider:
$$\begin{cases} d \cdot f_{out} = f_{in} \\ a_1 \cdot 1 + a_2 \cdot 2 + a_3 \cdot 4 + a_4 \cdot 8 = d \\ a_1 + a_2 + a_3 + a_4 = 1 \\ a_1, a_2, a_3, a_4 \in \{0, 1\} \end{cases}$$

---
[2] All unknowns and all parameters have positive integer values, except frequencies, which are real.

for dividers with a specified discrete set of possible values (e.g., $\{1, 2, 4, 8\}$). $f_{in}$ and $f_{out}$ are the input and respectively, output frequencies, $d$ is the division register and $a_1$, $a_2$, $a_3$, $a_4$ are parameters.

$$\begin{cases} f_{in} = d \cdot f_{out} \\ 1 \leq d \leq 65536 \end{cases}$$

for dividers with a specified interval of possible values (e.g., integers from 1 to 65536). $f_{in}$ and $f_{out}$ are the input and respectively, output frequencies, $d$ is the division register.

Using the equations we derived for all identified types of registers (the ones which have an influence on or are influenced by clock frequency) and some examples of user or application imposed constraints (Section 5) we were able to solve the resulting system for all unknowns by minimising the clock frequency.

Our system of equations falls into a category of particularly hard global optimisation problems, i.e. the so-called Mixed-Integer Nonlinear Programming (MINLP) problems. More details on this type of problems and on existing solvers will be given in Section 4.

### 3.3 Frequency Optimisation Graph

Combining the two approaches, the third hardware description scheme is the one we eventually used in the software we developed. This description presents the part of the hardware that is relevant to our study, under the form of a directed connected acyclic graph, in which source vertices are clock sources and sink vertices are usually dividers.

The graph contains the same hardware registers as the variables in the nonlinear system of equations. However, the frequency optimisation graph allows not only to find a global minimum for the frequency, but also to find all possible solutions, given a pre-imposed set of constraints. This is very useful, since the minimum may not always be the preferred solution.

Similarly to the first representation, registers are vertices and clocks are edges. We use only the registers modelled in the second representation. One additional type of vertex is needed to keep the structure a graph: the repeater, which replicates the input edge into as many output edges as necessary (the same clock goes into multiple blocks). Without this new type of vertex the structure would be a hypergraph, which is much more difficult to handle. Appendix D shows an example of such a graph, containing information from the basic clock module and the timers blocks of the TI MSP430.

The frequency optimisation graph is also an annotated graph, meaning that nodes and edges are annotated with extra information. Each type of node has specific information, e.g. possible frequencies for clock sources, division range or set for dividers, association between value of the selector and the selected clock for selectors etc. This data structure is then used by analysing and combining in an ideal way the information from both the graph structure and the annotations.

To gather information on the microcontroller, we used an XML file, which respects a slightly different grammar than the one presented in Appendix C. Slight modifications were made to node types, in order to accommodate for annotations. Concerning the parsing of the resulting XML file we used the same parser as the one used for the hardware dependency graph (Xerces C++ SAX parser). The parsed information is then converted into a graph data structure that can be easily manipulated by software, using the Boost Graph Library [5]. It this structure that will eventually be used in our tool for computing possible configurations.

### 3.4 Summary

To conclude this chapter, we may say that the frequency optimisation graph is our best chance to obtain possible configurations from the combination of hardware capabilities and user and/ or application constraints. We have also presented two other ways of representing hardware resources, which were very useful in developing the final form.

Representing hardware resources in a software useable manner will offer complete control over the hardware platform to software. Consequently, it will also enable the latter to reduce the power consumption of the former as a whole. This step is in line with our holistic approach to power optimisation, which aims at complementing existing schemes which only address parts of the systems, e.g. power-aware communication protocols.

As shown in the introduction of the report, the hardware description will be fed into the configuration computing tool, which will generate optimal configurations for the given requirements. This will enable the application to have a small set of configurations, each with its own clock frequency, at compile time and will allow it to switch among them at will. Once the possible configurations are calculated for each application, the programmer can include code in the application or in the operating system (e.g., under the form of a service), that will switch from one hardware configuration to another (and at the same time from one clock frequency to another), in function of the application's needs.

# 4 Connecting Hardware and Software

In the previous chapter, different ways of representing hardware resources have been presented. This chapter will describe the operation of the hardware configurations computing tool from Figure 1. The tool is dedicated to using the above representations to configure and allocate hardware resources such that user and/or application requirements are respected and that power drawn from the battery is minimum. As pointed out above, minimising power drawn from the battery will increase its capacity.

In making hardware and software work together in an optimal manner, we worked through several steps, illustrated in Figure 4 (only green items have been addressed in this study). First, we developed a novel timer allocation algorithm, since timers are one of the key microcontroller subsystems in reducing operating frequency. We then used this allocation scheme to place constraints on hardware timers, which allow us to obtain all valid hardware configurations, by simply walking through the frequency optimisation graph and applying the constraints imposed by each vertex to the set of solutions. Both schemes are described in more detail in the following sections.

## 4.1 Software Timer Allocation

The allocation of software timers to hardware timers is an important factor in determining the minimum frequency at which the microcontroller can operate. Despite the importance of the matter, popular sensor network operating systems do not consider this problem in their studies. As we saw in Subsection 2.3, the OSs assign all software timers to one clock or hardware timer. Sometimes, the clock or hardware timer also has the role of OS clock. The frequency is often very high (e.g., 4 MHz in TinyOS 2.x), as compared to its optimal value, in order to accommodate a decent timer resolution. Our idea was to create an allocation scheme that will calculate the minimum frequency required to provide all application and OS timers, while spreading them throughout the available hardware timers.

If the number of software timers is less than or equal to the number of hardware timers, then each software timer can be assigned to one hardware timer. However, if there are more software timers than hardware timers, as is almost always the case, a strategy is needed in order to find the best distribution. First, it is important to notice that a hardware timer ticking every $h$ time units can not only accommodate a software timer with value $h$, but also timers with values multiple of $h$, e.g. $2h, 3h$ etc, at no additional cost in power drawn from the battery. This is possible by having the hardware timer tick every $h$ time units and having the timers multiple of $h$ managed by software.

Therefore, a good first step in the timer allocation scheme is to separate the software timer set into small sets of timers which are multiple of another timer, as shown in Algorithm 1.

Once the timers are separated into sets of multiples, the problem is reduced to allocating the timer ticking at a maximum value of time units from each set of multiples. Once again, if the number of software timers is less than or equal to the number of hardware timers, each software timer can simply be served by one hardware timer. Otherwise, a clustering scheme is necessary. In order for the clustering to be optimal, the frequency of the hardware timer that will accommodate the cluster must be minimised.

If a hardware timer must serve a cluster of software timers which are not multiples of one another, then the minimum operating frequency is given by the multiplicative inverse of the greatest common divisor (GCD) of the time intervals required by software. Hence, to minimise operating frequency, we need to maximise these values (seeing that $frequency = \frac{1}{timeperiod}$), the GCD of each cluster, when forming the partition.

In order to obtain a function whose optimisation will maximise the GCDs, we will consider the problem in terms of number of interruptions. For each hardware timer, the lower the frequency, the smaller the number of interruptions per time unit and thus, the less energy consumed. So, if we want to have a global minimum, all we have to do is minimise the sum of all numbers of interruptions per time unit, which boils down to minimising the sum of all hardware timer frequencies, as shown in Equation 3 (where $sw$ is the set of software timers ticking at a maximum value of time units from each set of multiples, $\mathcal{P}(sw)$ is the partition to evaluate and $C_i$ are the clusters).

$$W(\mathcal{P}(sw)) = \sum_{i=1}^{|hw|} \frac{1}{GCD(C_i)} \qquad (3)$$

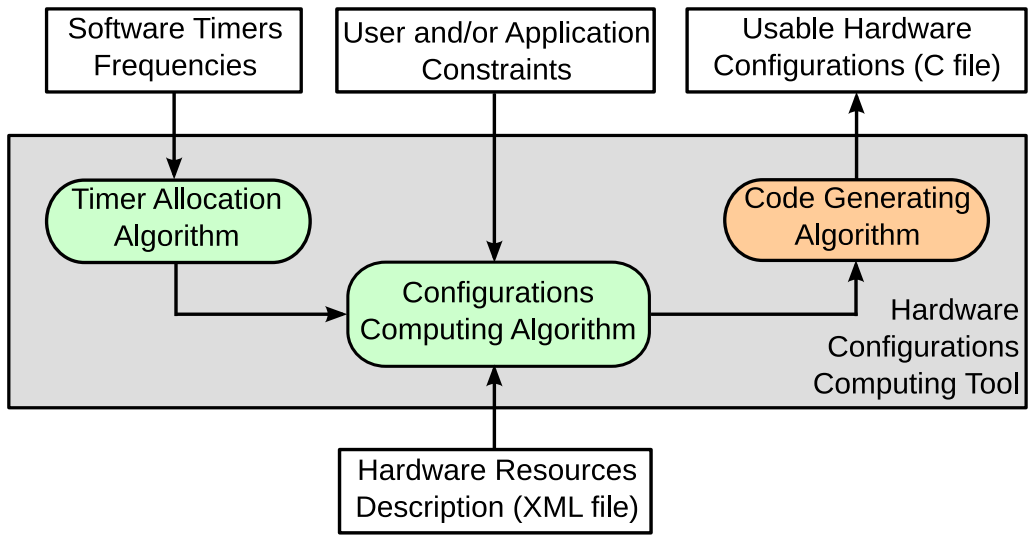

Figure 4: Detail of the hardware configurations computing tool

This is a set partitioning NP-complete optimisation problem. Jensen's algorithm [22, 13] is a dynamic programming clustering algorithm which fits our problem very well. This algorithm considers the problem of partitioning a set of pre-defined cardinality or size (number of software timers) into a pre-defined number of clusters (i.e. hardware timers). The author proves that it always converges on the best clustering solution.

In order for the algorithm to work, the function to be optimised must be additive and this condition is fulfilled by our function. Although it is a huge improvement over total enumeration, Jensen's algorithm still needs to store the value of the function $W$ for almost all subsets of $sw$, which makes it unusable for large problems (i.e. it is limited to 17 elements on a 2 GHz Pentium M with a 1 GB of RAM). However, the problem we are dealing with is reasonably small, which allows us to successfully use Jensen's algorithm.

A brief description of how we applied Jensen's algorithm to our problem follows. The goal of the scheme is to build a graph very similar to the Hasse diagram ([30]) of a family of subsets over the set to be partitioned, i.e. $sw$, partially ordered by inclusion. The only difference with the Hasse diagram is that in Jensen's graph, one subset can have multiple occurrences at different *stages* of the graph.

$$mn(s) = \begin{cases} s \cdot \frac{|sw|}{|hw|} & \text{if } |sw| \propto |hw| \\ \begin{cases} (\frac{|sw|}{|hw|} + 1) \cdot s & \text{if } 1 \leq s \leq |sw| - |hw| \cdot \frac{|sw|}{|hw|} \\ |sw| - (|hw| - s) \cdot \frac{|sw|}{|hw|} & \text{if } |sw| - |hw| \cdot \frac{|sw|}{|hw|} < s \leq |hw| \end{cases} & \text{otherwise} \end{cases} \quad (4)$$

$$mx(s) = |sw| - |hw| + s \quad (5)$$

In-clearer, Jensen's structure is a directed acyclic graph whereof the source is the empty set $\emptyset$ and the sink is $sw$ itself. Between the source and the sink, the graph is composed of a certain number of *stages*, calculated as shown in Equation 6. There are no edges between vertices of the same *stage*. Edges are only possible between vertices of adjacent *stages* and only if a relation of inclusion exists between the two.

$$s = \begin{cases} |hw| & \text{if } |sw| \geq 2 \cdot |hw| \\ |sw| - |hw| & \text{otherwise} \end{cases} \quad (6)$$

For the construction of each *stage*, the scheme computes the minimum and the maximum cardinalities of subsets of that *stage*, as shown in Equations 4 and 5. The equations come from Jensen's observation that it is possible to eliminate redundant partitions, by considering distribution forms. A distribution form is a subset of partitions sharing the same format. For example, if we want to break up a set of 4 elements into two subsets, then {3}-{1} (three elements in the first subset and one element in the second) and {2}-{2} (two elements in each subset) are the only possible distribution forms.

**Algorithm 1** Separation of timers in sets of multiples
1: $sw$ set of software timers
2: $i, j, m$ positive integers
3: $c$ table of sets of multiples
4:
5: $i \leftarrow 0$
6: **while** $sw \neq \emptyset$ **do**
7: $\quad m \leftarrow min(sw)$
8: $\quad c[i] \leftarrow \emptyset$
9: $\quad$ **for** $j = 0$ to $|sw| - 1$ **do**
10: $\quad\quad$ **if** $sw[j] \propto m$ **then**
11: $\quad\quad\quad c[i] \leftarrow c[i] \cup \{sw[j]\}$
12: $\quad\quad\quad sw \leftarrow sw \setminus \{sw[j]\}$
13: $\quad\quad$ **end if**
14: $\quad$ **end for**
15: $\quad i \leftarrow i + 1$
16: **end while**

Once the cardinalities are computed, all subsets of $sw$, with cardinalities comprised within these limits are added to the graph at the corresponding *stage*. Edges are also added based on inclusion of the subsets from the previous *stage*. For our problem, the weight of an edge is the multiplicative inverse of the GCD of the elements contained in the difference of the two adjacent subsets.

The construction of the graph as applied to our problem is illustrated in Algorithm 2 and Jensen's numerical example is reproduced in Appendix E. Once the graph is created, all we have to do to obtain the optimal partition is to apply Dijkstra's shortest path algorithm (from the Boost Graph Library [5]) to the graph.

To finish this section, Jensen's algorithm allows us to assign software timers to hardware timers in an optimal fashion. Thus, we obtain user and application constraints (assuming the user or application included their timers in $sw$) under the form of constraints on hardware registers of Timer A and/or Timer B of the TI MSP430. During our study, we used these constraints both in resolving our nonlinear system of equations (Subsection 3.2) and in computing hardware configurations as we will see in the next section.

## 4.2 Computation of Configurations

Frequency scaling implies a lot of configuration changes if we want to keep continue fulfilling user and application requirements. This is why a hardware configuration management tool is essential for our project. Our initial idea, described in Subsection 3.1 summarised very well the goal: being able to enumerate possible hardware configurations and to choose the one that best fits our scenario. However, total enumeration is huge task which is practically infeasible.

### 4.2.1 Solving the MINLP Problem

This is why we turned our attention to analytical methods, like expressing hardware capabilities as equations, as shown in Subsection 3.2. This method was very successful, especially in combination with the Jensen's algorithm [13], for the derivation of user and application constraints. Other constraints come from oscillator capabilities, as mentioned in Subsection 3.2.

Appendix F shows an example of the nonlinear system of equations for TI MSP430's basic clock module and timer A. Constraints on timer A's registers are drawn from Jensen's algorithm and constraints on clock frequencies come from the TI MSP430 user guide [12]. The system is a MINLP problem, i.e. a Nonlinear Programming (NLP) problem in which some of the variables are required to take integer values. A NLP problem is defined as follows [9]: there is one scalar-valued function $F$, of several variables, that we seek to minimise or maximise subject to one or more other such functions that serve to limit or define the values of these variables. $F$ is known as the "objective function", while the other functions are called the "constraints".

**Algorithm 2** Jensen's clustering algorithm applied to our problem [13]
 1: $sw$ set of software timers
 2: $hw$ set of hardware timers
 3: $g$ Jensen's graph
 4: $i, j, s$ positive integers
 5: $first, last, set$ sets
 6:
 7: $g \leftarrow g + vertex(\emptyset)$
 8: **for** $i = 0$ to $s$ **do**
 9: $\quad j \leftarrow mn(s)$
10: $\quad$ **repeat**
11: $\quad\quad first \leftarrow$ first set of $j$ elements
12: $\quad\quad last \leftarrow$ last set of $j$ elements
13: $\quad\quad set \leftarrow first$
14: $\quad\quad$ **repeat**
15: $\quad\quad\quad g \leftarrow g + vertex(set)$
16: $\quad\quad\quad g \leftarrow g + edges(set)$
17: $\quad\quad\quad set \leftarrow$ next set of $j$ elements
18: $\quad\quad$ **until** $set > last$
19: $\quad\quad j \leftarrow j + 1$
20: $\quad$ **until** $j > mx(s)$
21: **end for**
22: **if** $mx(s) < |sw|$ **then**
23: $\quad g \leftarrow g + vertex(sw)$
24: $\quad g \leftarrow g + edges(sw)$
25: **end if**
26: Dijkstra shortest path in $g$ from $vertex(\emptyset)$ to $vertex(sw)$

Software for this particularly difficult kind of global optimisation has followed two approaches: outer approximation and branch and bound. The BARON solver [23] is based on the second approach and is available for online use on the NEOS server for optimisation [28]. The server requires models written in the GAMS modeling language [4]. Using this online solver allowed us to check that our system is solvable and gave us an idea on the approach to take in calculating all possible hardware configurations for a given set of constraints.

#### 4.2.2 Walking the Graph

This is how we came to use the hardware description presented in Subsection 3.3 and illustrated in Figure 5.

Using this annotated graph, we calculate possible hardware configurations in the following way: we use the Boost [5] implementation of the depth-first visit algorithm to walk the graph in post-order. This means that we step through the items of the graph, by means of the connections between parents and children and that we traverse children before their respective parents.

This allows us to start with the sink nodes and work our way up to the source nodes (which are all clock sources), while adding more and more constraints to the clock frequency on the way. When the walk is finished, we obtain a list of possible clock frequencies and the associated microcontroller configurations. Constraints can be easily added or removed by accessing the graph structure.

The traversing operation is evidently different for each type of vertex. Here are some examples:

- **selector**: copy all child configurations for each selection option and add that option to each configuration.
- **divider**: for all child configurations, multiply the clock frequency by the value of the divider and add that value to the configuration. If the node has no children, but has a constraint, create new configurations with all possible values of the divider, given the constraint. If the node has no children and no constraint, do not create any configuration, since the node imposes no constraint on its parents.

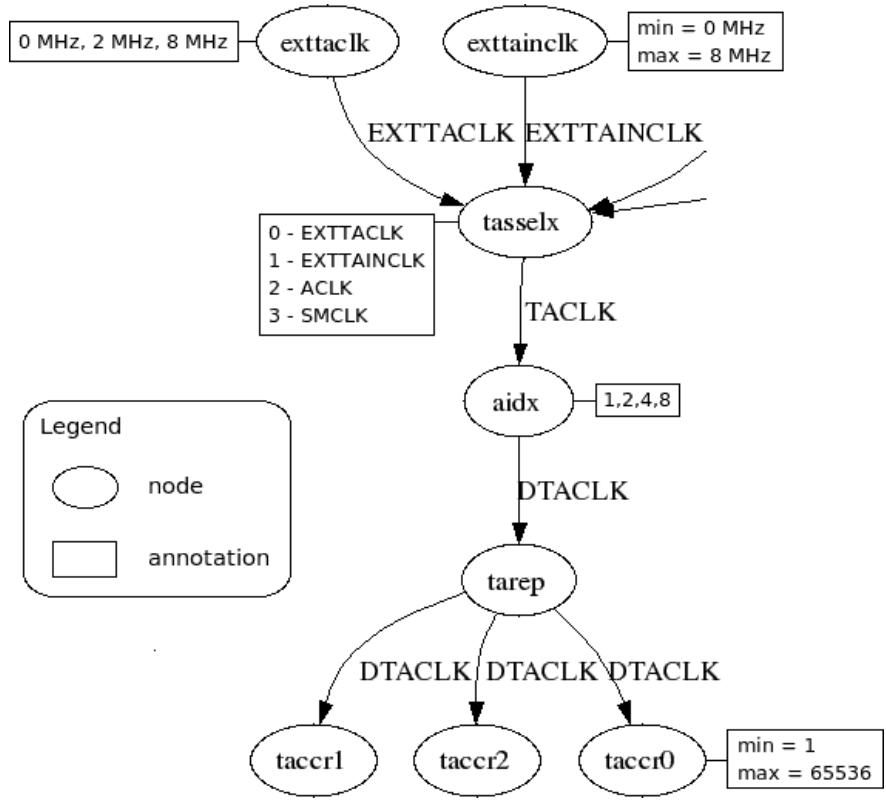

Figure 5: Annotated frequency optimisation graph

- **repeater**: compute the intersection of all the sets of hardware configurations coming from the children. The result of the intersection is a set of configurations obtained by merging child configurations for which all common registers have the same values and by discarding all other configurations.

For performance reasons, it is useful to compute an upper limit for clock frequencies (using the constraints) before performing the graph walk. This will significantly reduce the number of hardware configurations to be considered, from the very beginning.

## 4.3 Summary

This chapter presented the two main features of our power saving scheme: enhanced hardware resource allocation and application constraints-compliant configurations. Their effectiveness is suggested by the mathematical equations of power and energy versus frequency and also by battery discharge models.

These two schemes are the heart of our holistic approach to the management of power consumption. The combination of the two takes full advantage of the new perspectives offered by the detailed knowledge of available resources. It can optimise the use of these resources and thus energy on all levels of the system, from the application level down to the OS level and through the communication protocols, as we will see in the next chapter.

This approach based on frequency scaling, used in combination with LPMs, energy-aware communication protocols and software is likely to save large amounts of power and contribute to increasing the battery capacity by lowering the peak power drawn, as discussed in 2.4.

# 5 Towards Energy-aware Software

Reliable and accurate information for solving the excessive power consumption problem can only be obtained from applications themselves. Application-directed frequency scaling offers the opportunity to use just the right hardware resources instead of a general purpose estimate used in operating systems.

However, it requires the application to have a clear representation of its needs at all times. The needs of the application are also very important in an improved hardware resources allocation scheme, since its role is to assign hardware to fulfil software requirements. We will therefore take an example of WSN software and try to derive its requirements related to timing, which is the main issue for frequency scaling and for hardware timers.

The requirements will then be input to the hardware configurations computation tool. We compare the results from each step of the approach to common algorithms for this type of problem, e.g. the greedy algorithm. Then we discuss the outcome and we try to argue for the effectiveness of our approach.

## 5.1 Case Study: ZigBee [19, 25]

ZigBee is the name of a specification for a suite of high level communication protocols, targeted at RF applications that require a low data rate, long battery life, and secure networking. It is based on the IEEE 802.15.4 standard and the relationship between the two is the same as that between IEEE 802.11 and Wi-Fi. In our study we used the most recent version of the specification which dates from December 2006 [19].

ZigBee specifies an entire communication stack: IEEE 802.15.4's physical and MAC layers, network layer using the AODV routing algorithm and application layer. The reason why we chose to study ZigBee is its becoming a wide spread standard in WSNs and ZigBee-compliant chips are already on sale. In addition to that, it is a complex set of communication protocols which are likely to provide interesting constraints for us to test, especially as far as timers go.

In the following sections we will study each of ZigBee's layers, from the point of view of their frequency and timer needs. We will start with the physical layer and work our way up to the application layer.

### 5.1.1 Physical Layer

The physical layer of a protocol stack provides the means of transmitting raw bits over a physical data link connecting network nodes. The bit stream may be grouped into code words or *symbols*, and converted to a physical signal, that is transmitted over a physical transmission medium. The shapes of the electrical connectors, which frequencies to broadcast on, what modulation scheme to use and similar low-level parameters are specified here.

The study of the physical layer might seem irrelevant to our work, since we are concerned with software. However, time intervals are specified in number of *symbols* in ZigBee's MAC layer. As explained below, modulation schemes are the ones determining the *symbol* duration and the number of bits per *symbol* (or of *symbols* per octet). It is therefore essential to study the different modulation schemes available in ZigBee and listed in Table 2[3].

Naturally, ZigBee uses digital modulation schemes [27]. In digital modulation, an analog carrier signal is modulated by a digital bit stream. The changes in the carrier signal are chosen from a finite number of alternative *symbols* (the modulation alphabet).

The characteristics of the analog signal, phase, frequency, amplitude or combinations of them, are used to represent the digital bit stream. Each of these phases, frequencies or amplitudes are assigned a unique pattern of binary bits. Usually, each phase, frequency or amplitude encodes an equal number of bits. The fixed-size set of bits comprises the *symbol* that is represented by the particular characteristic (phase, frequency etc).

If the alphabet consists of $M = 2^N$ alternative *symbols*, each *symbol* represents a message consisting of $N$ bits. If the *symbol* rate (also known as the baud rate) is $f_S$ *symbols*/second, the data rate is $N \cdot f_S$ bit/second. For example, with an alphabet consisting of 16 alternative *symbols*, each *symbol* represents 4 bit and the data rate is four times the baud rate.

---

[3]Data is reproduced from [25].

| PHY (Mhz) | Frequency band (Mhz) | Data parameters | | Symbol duration (us) | SHR duration (symbols) | Symbols per octet |
|---|---|---|---|---|---|---|
| | | Bit rate (kb/s) | Symbol rate (ksymbol/s) | | | |
| 868/915 | 868-868,6 | 20 | 20 | 50 | 40 | 8 |
| | 902-928 | 40 | 40 | 25 | 40 | 8 |
| 868/915 (optional) | 868-868,6 | 250 | 12.5 | 80 | 3 | 0.4 |
| | 902-928 | 250 | 50 | 20 | 7 | 1.6 |
| 868/915 (optional) | 868-868,6 | 100 | 25 | 40 | 10 | 2 |
| | 902-928 | 250 | 62.5 | 16 | 10 | 2 |
| 2450 | 2400-2483,5 | 250 | 62.5 | 16 | 10 | 2 |

Table 2: IEEE 802.15.4 physical layer characteristics

Therefore, and as shown in Table 2, it is easy to derive *symbol* duration in seconds and the number of bits per *symbol* from bit and baud rates, which are usually given in specifications. For the purpose of our study, we decided to use the data given for the 2400-2483,5 MHz frequency band, since it is the most commonly employed for amateur transmissions. Our symbol duration will consequently be 16 $\mu$s. This will be useful in converting MAC layer time intervals into seconds, as we will see below.

### 5.1.2  MAC Layer

Zigbee's MAC sublayer provides two services: the data service and the management service. The former enables the transmission and reception of MAC protocol data units across the physical layer. The latter allows the transport of management commands to and from the next higher layer.

The standard allows the use of an optional superframe structure, whereof the format is defined by the so-called ZigBee coordinator. The superframe is bounded by network beacons sent by the coordinator and is divided into 16 equally sized slots. The beacon is transmitted in the first slot of each superframe. The superframe is used for synchronisation and for network identification. The structure of the superframe is illustrated in Figure 6.

The superframe consists of a Contention Access Period (CAP) and of an optional Contention Free Period (CFP). The CAP is the time duration in *symbols* during which the devices can compete with each other to access the channel using CSMA-CA and transmit the data. CFP is the time duration for which certain low-latency application devices are given exclusive rights over the channel and the devices can directly start transmitting the data. There can be as many as 7 slots assigned for CFP transmissions. These transmissions start immediately after the CAP. The inactive period is optional and during this period the coordinator goes to a power save mode. Therefore, during this time, there will be no beacon transmissions and the other devices also go to sleep mode for this duration.

After this description we cannot help but notice the leitmotiv of ZigBee's MAC layer is time synchronisation, which means that the timer service will be very demanded. During the study of this layer, we identified no less than 17 time intervals (for the beacon-enabled mode), among which we can name the Short and Long Interframe Spaces (SIFS and LIFS), the superframe duration, the time to wait for an acknowledgement, the backoff period for the CSMA-CA algorithm etc. All time intervals are expressed in *symbols* and must be multiplied by 16 $\mu$s (since we chose the 2400-2483,5 MHz frequency band) to convert them into veritable time intervals.

---

[1]Figure is available without copyright from [15].

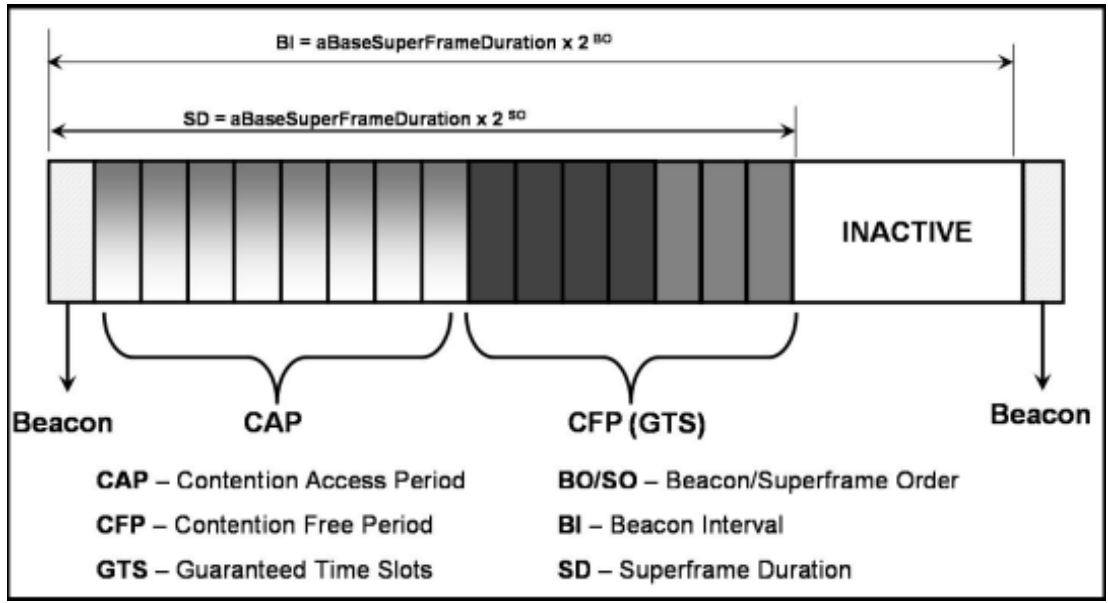

Figure 6: ZigBee superframe structure[1]

### 5.1.3 Network Layer

The routing algorithm used by ZigBee follows a hierarchical strategy with table-driven optimisations applied where possible. It is a combination of the well-studied public-domain AODV algorithm [20] and Motorola's cluster-tree algorithm [2]. ZigBee's routing algorithm allows various network organisations: cluster of clusters, single cluster, mesh. Brief descriptions of both algorithms follow. We will conclude the section with an overview of the algorithm's timing needs.

**AODV Algorithm** AODV is an on-demand route construction algorithm, which means that it builds routes between nodes only as desired by source nodes and it maintains them only for as long the sources need them. The algorithm makes use of sequence numbers to ensure route freshness and it is loop free.

Routes are built using a route request – route reply query cycle. When a source node desires a route to a destination for which it does not already have a route, it broadcasts a route request (RREQ) packet across the network. Nodes receiving this packet update their information for the source node and set up backwards pointers to the source node in the route tables. A node receiving the RREQ may send a route reply (RREP) if it is either the destination or if it has a route to the destination with corresponding sequence number greater than or equal to that contained in the RREQ.

As the RREP propagates back to the source, nodes set up forward pointers to the destination. Once the source node receives the RREP, it may begin to forward data packets to the destination. Once the source stops sending data packets, the route is considered inactive, the links will time out and eventually be deleted from the intermediate node routing tables.

**Cluster-Tree Algorithm** The cluster-tree protocol spans the logical link and the network layers and uses link-state packets to form either a single cluster network or a potentially larger cluster tree network (or multi-cluster network). Nodes select a cluster head and form a cluster in a self-organised manner. In the cluster formation process the cluster head assigns a unique node ID to each member node. Self-developed clusters connect each other using the Designated Device. The Designated Device is a special node that has high computing ability and large memory space; in most applications it is also the gateway between the network and the Internet. The Designated Device assigns a unique cluster ID to each cluster.

ZigBee's routing layer requirements for timers are not as demanding as the ones coming from the MAC layer. However, we could still find some ten time intervals, like the route discovery time, the RREQ retry interval, the duration of an association permit delivered by a coordinator etc. Some of these intervals are still expressed in symbols or superframe durations, therefore conversion is needed.

### 5.1.4 Application Layer

The ZigBee application layer consists of the application support sub-layer (APS), the ZigBee device object (ZDO) and the user- or manufacturer-defined application objects.

The responsibilities of the APS sub-layer include maintaining tables for binding, which is the ability to match two devices together based on their services and their needs, and forwarding messages between bound devices. Another responsibility of the APS sub-layer is discovery, which is the ability to determine which other devices are operating in the personal operating space of a device.

The responsibilities of the ZDO include defining the role of the device within the network: ZigBee coordinator or end device, initiating and/or responding to binding requests and establishing a secure relationship between network devices.

The user- or manufacturer-defined application objects implement the actual applications according to the ZigBee-defined application descriptions.

The application layer in itself has very little timing requirements. Nevertheless, it contains application objects, which might have timing requirements of their own. In function of the applications, this may considerably increase the demand for timers.

### 5.1.5 Analysis of Overall Requirements

After having identified timing requirements for each layer, we must consider the entire protocol stack. Putting everything together showed that the ZigBee protocol stack needs some thirty software timers, both periodic and one shot. These timers range from 128 $\mu$s (the time to scan a channel for energy detection at the MAC layer) to 15.36 seconds (also at the MAC layer).

This list of software timers must be transmitted to the resource allocation software, which will first create a partition of the initial set, in which for each subset, all elements are multiple of the minimum of the subset. Table 3 shows the result of the algorithm as applied to ZigBee timers.

| Minimum Timer ($\mu$s) | Other Timers ($\mu$s) |
| --- | --- |
| 128 | 640, 15 360, 30 720, 122 880, 768, 491 520, 15 360 000, 1 280 000, 10 000 000, 64 000, 1 600 000 |
| 192 | 960, 1 344 |
| 320 | 2 240, 1 000 000 |
| 1000 | 254 000, 3 000, 9 000 |

Table 3: ZigBee timers separated in sets of multiples

Once the list of software timers has been reduced to only the set of minimums of each subset, all that is left to do is to assign hardware timers to the reduced set. For the rest of our discussion, let us suppose we use a microcontroller with two hardware timers. Therefore, we have to find a two-cluster partition of our reduced set of timers, which minimises the sum of hardware frequencies. To that end, we apply Jensen's algorithm, described in Subsection 4.1. The intermediate steps and the results of the scheme are presented in Table 4.

| Minimum Timer ($\mu$s) | PF[1] ($\mu$s) | GCD ($\mu$s) | Hardware Timer (kHz) |
|---|---|---|---|
| 128 | $2^7$ | 64 | 15,625 |
| 192 | $2^6 \cdot 3$ | | |
| 320 | $2^6 \cdot 5$ | | |
| 1000 | $2^3 \cdot 5^3$ | 1000 | 1 |

Table 4: Results of Jensen's algorithm on ZigBee timers

In order to understand the extent of the improvement brought by our approach over ordinary methods we could have applied and over common WSN operating systems, let us consider three alternatives to our scheme:

- ***A greedy algorithm.*** This is any algorithm that follows the problem solving metaheuristic of making the locally optimum choice at each stage with the hope of finding the global optimum. In our case, it boils down to iteratively choosing the maximum software timer from the reduced set and assigning it to the hardware timer with the smallest current frequency. Thus, a greedy algorithm makes one greedy choice after another, reducing each given problem into a smaller one. In other words, a greedy algorithm never reconsiders its choices. This is the main difference from dynamic programming (e.g., Jensen's algorithm), which is exhaustive and is guaranteed to find the solution.

  ***Result:*** First hardware timer runs at 15,625 kHz and provides for 320 $\mu$s and 192 $\mu$s. Second hardware timer is set at 125 kHz and provides for 1000 $\mu$s and 128 $\mu$s.

- ***Calculating the GCD of the whole set.*** In our scheme, we try to minimise the hardware frequency for each cluster of software timers. However, if we wanted to keep the timer virtualisation used in most WSN OSs (i.e., use only one hardware timer for all software timers), we could still improve it simply by setting the hardware timer frequency to the multiplicative inverse of the GCD of the set of software timers.

  ***Result:*** The hardware timer is set at 125 kHz and provides for all software timers.

- ***Keeping the approach of current WSN OSs.*** This means we maintain the fixed frequency and we only use LPMs for saving energy.

  ***Result:*** The hardware timer runs at 4 MHz and provides for all software timers.

Figure 7(a) shows a comparison of the timer frequency resulting from each of the four approaches. Figure 7(b) zooms in on the first three approaches (since they are markedly better than the last one), giving us a better idea of the differences among them.

Jensen's algorithm performs very well as far as minimising the frequencies of hardware timers and using available resources in an optimal way.

The greedy algorithm produces a suboptimal solution. This is expected, since our problem does not have two essential properties which allow the greedy algorithm to generate good solutions. First, our problem does not have an optimal substructure, meaning that the optimal solution does not necessarily contain optimal solutions to subproblems. And second, our problems does not have the greedy choice property, meaning that the choice made by the greedy algorithm may depend on choices made so far, but must not depend on future choices or on all the solutions to the subproblem.

The solution offered by the third algorithm is better than the one from the greedy algorithm. It uses only one of timer running at one of the frequencies produced by the greedy algorithm.

The scheme currently employed in WSN operating systems is the worst among the four.

In a nutshell, our hardware timer allocation algorithm is very effective in reducing the operating frequency. Moreover, its execution time is reasonably low for a set of up to 17 software timers (which are not multiple of each other). This allows a very large extended set of software timers and should be more than enough for sensor applications. The allocation algorithm is also essential for the frequency scaling scheme, which uses the configurations computation tool.

---
[1] Prime Factorisation

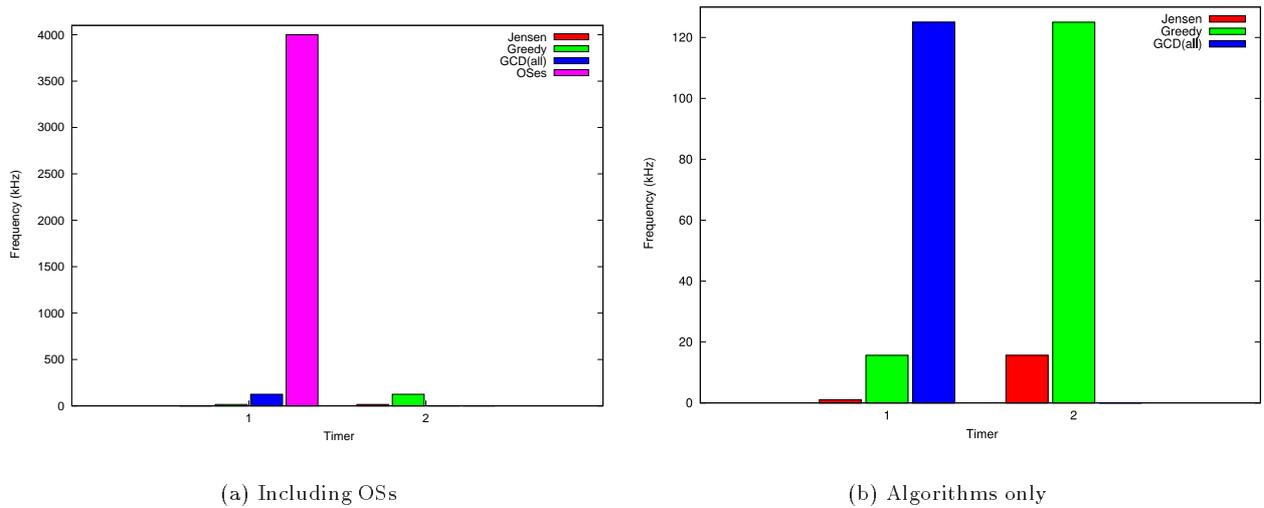

(a) Including OSs         (b) Algorithms only

Figure 7: Comparisons of hardware timer frequencies

The analysis of battery discharge behaviour and of power and energy equations suggest that the above mentioned tool will bring significant improvements to the power consumption in sensors. Moreover, it will allow the use of voltage scaling, as soon as it becomes available for the microcontrollers in use. The experimental evaluation of the effectiveness of this tool is left as future work.

## 5.2 Summary

In conclusion, this section studied a very famous WSN communication stack, which is likely to become a generally accepted standard, i.e. ZigBee. A brief presentation of each of ZigBee's layer was given, in order to obtain a better understanding of those parts of the protocol stack which are of interest to our study. We analysed ZigBee's needs in terms of timers and we determined that our holistic approach to energy saving can bring significant improvement even to those parts of a sensor network which are already power-aware, in this case power aware communication protocols.

# 6 Conclusion

In this report, a holistic approach to power consumption management on sensor platforms was introduced. This solution is based on the observation that available WSN operating systems offer little alternatives to and models of active administration of the energy consumption. Although LPMs and energy-aware communication bring a certain amount of control, currently employed techniques are not fully developed. No system comprises an infrastructure considering the energy consumption at the hardware platform level (CPU and peripherals). This would allow the establishment of application-driven policies for the management of power in WSNs.

Even though the proposed solution is not entirely implemented nor fully tested at this time, this report addresses the bulk of the work and lays the groundwork for the remaining investigations.

## 6.1 Contributions

Our work brings two complementary methods to reduce both the power drawn from the sensor's battery and the power dissipation within the microcontroller. Their goal is to facilitate the collaboration between a very resourceful hardware platform and the user or the application.

A frequency-optimal algorithm is presented that matches software timers required by higher layers to available hardware timers. In addition to finding the minimum operating frequency for a given set of timers, it will enable the application framework to recalculate a new optimal timer frequency at any time.

Frequency scaling can be applied at the hardware platform level only if all dependencies are considered. These include the clock choices made by each peripheral, constraints on the clock generators and on peripheral outputs as well as several other details. To account for all the above mentioned dependencies, we developed a hardware configurations computation tool. The idea behind this approach is that full control over the whole microcontroller is needed if we are to respect all the relationships between hardware components at the time of the frequency scaling. This is why our tool uses a detailed description of the hardware platform to compute all possible configurations for a given set of constraints.

The application layer will therefore have the possibility to scale the frequency, in function of its current level of activity, without disturbing components which depended on the previous operating frequency. This action, when combined with LPMs, is thought to be very efficient at saving energy.

When completed, the infrastructure we introduced, will enable software to apply chosen power management policies and thus consume just the right amount of energy. A case study is presented which demonstrates the applicability and the benefits of the proposed approach.

## 6.2 Future Work

**Short term improvements.** The algorithm proposed for assigning hardware timers to requested software timers is strict as far as timer frequencies are concerned. Even though two timer periods may have very close values, they will both be considered for hardware timer allocation. However, it is frequent that applications do not need so much precision. Therefore, two near values may be merged into a single timer. The benefit of this action is that there will be less processing both in assigning timers and in managing them during operation.

**Longer term actions.** Several new developments and extensions to existing work must be accomplished. The former include: software timers description and transmittal to the configuration tool, development of a code generator that will transform computed configurations into executable code. As far as the latter is concerned, the efficiency of our configuration computation tool still needs to be confirmed experimentally.

Part of the work we have already carried out consisted of describing the hardware platform in a suitable way for the configuration tool to use. The same task must to be completed for software requirements and needs. These may include software timers, processing frequencies, transmission rates etc. The description must be practical and easily feedable to the above mentioned tool.

The configuration computing program generates a certain number of hardware configurations in the form of values to assign to hardware registers. Nevertheless, in order for the application to be able to use them, the configurations must take the form of pieces of code which will assign the computed values to respective registers. This is why a code generating tool must be developed.

As to experimental work, the generated configurations still need to be tested first in simulation tools and then in real sensing devices. The results of the tests are essential for the certification of theoretical results and of intuitions.

# Appendices

## A TinyOS Timer System (Subsection 2.3)

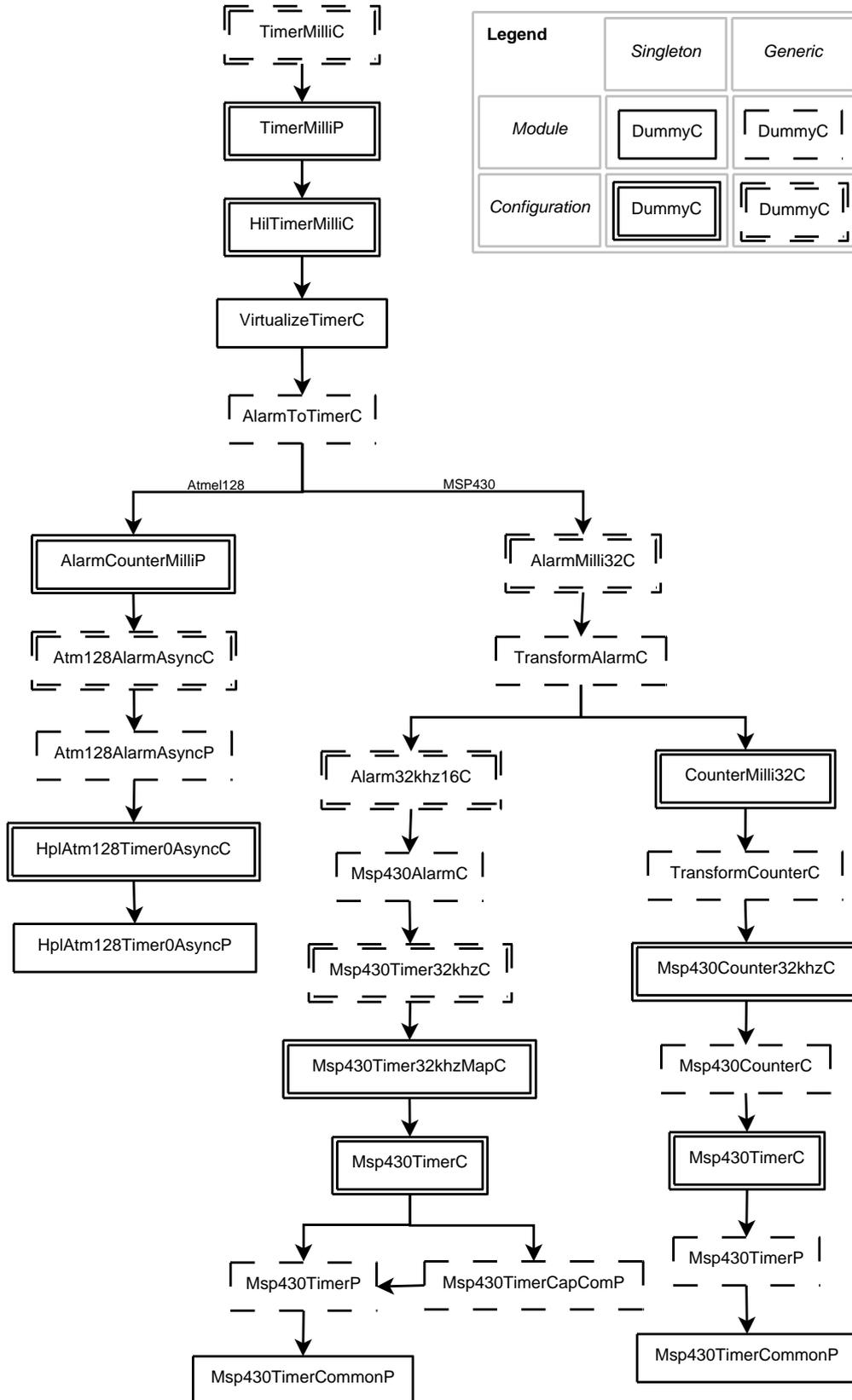

## B  Flash Dependency Graph (Subsection 3.1)

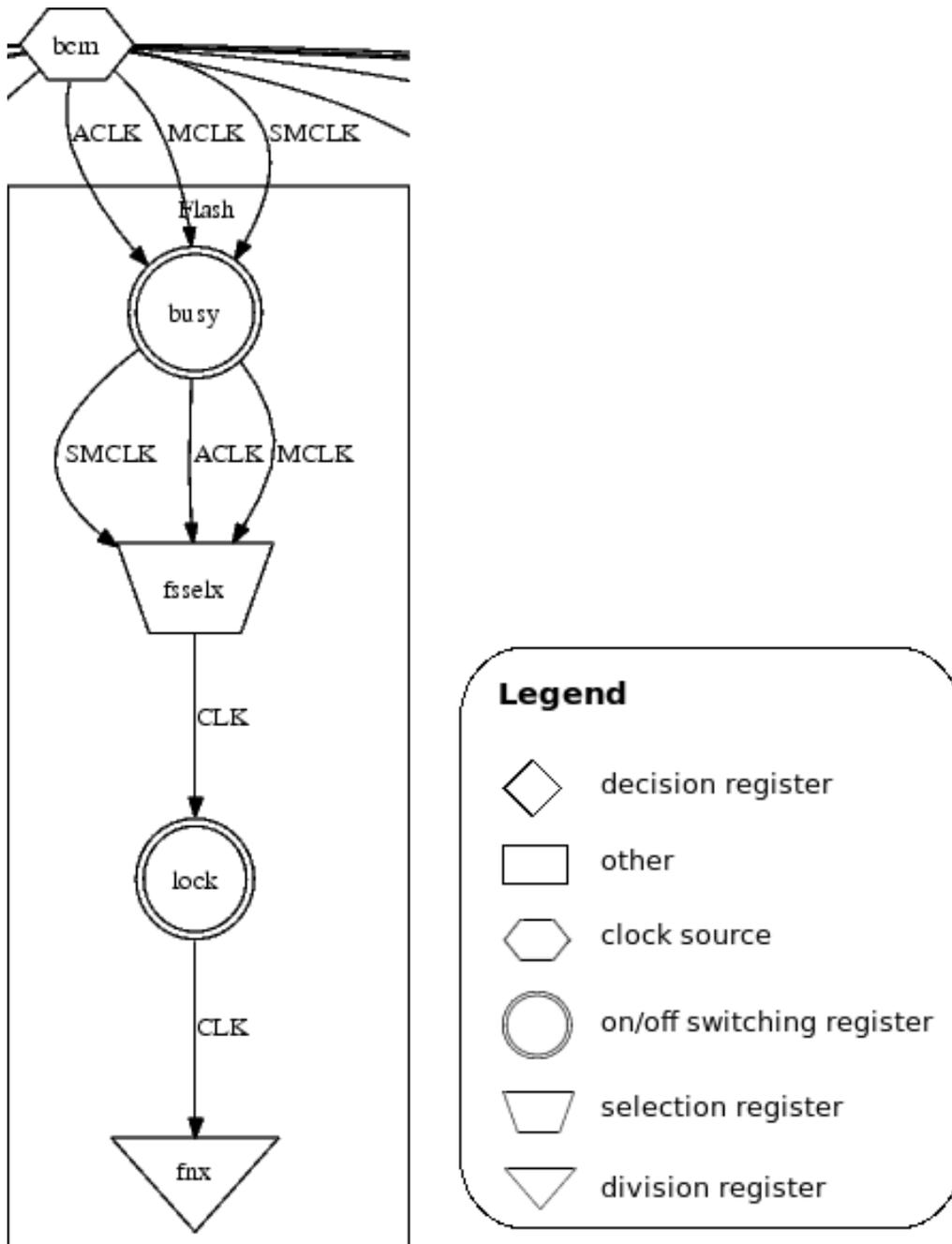

## C First XML Grammar (Subsection 3.1)

```xml
<?xml version = "1.0" encoding = "UTF-8"?>
<xsd:schema    xmlns:xsd = "http://www.w3.org/2001/XMLSchema"
            targetNamespace = "http://www.wasp-project.org/wsnfreq/deps"
            xmlns = "http://www.wasp-project.org/wsnfreq/deps"
            elementFormDefault = "qualified">
<!-- Hardware Dependency Graph Bucket -->
<xsd:element name = "hwDependencyGraphs">
    <xsd:complexType>
        <xsd:sequence>
            <xsd:element ref = "node" maxOccurs = "unbounded"/>
            <xsd:element ref = "hwDependencyGraph" maxOccurs = "unbounded"/>
        </xsd:sequence>
    </xsd:complexType>
</xsd:element>
<!-- Hardware Dependency Graph -->
<xsd:element name = "hwDependencyGraph" type = "hwDependencyGraphType"/>
<xsd:complexType name = "hwDependencyGraphType">
    <xsd:sequence>
        <xsd:element ref = "node" maxOccurs = "unbounded"/>
        <xsd:element ref = "edge" maxOccurs = "unbounded"/>
        <xsd:element ref = "hwDependencyGraph" minOccurs = "0" maxOccurs = "unbounded"/>
    </xsd:sequence>
    <xsd:attribute name = "id" type = "xsd:ID"/>
</xsd:complexType>
<!-- Node and Node Type -->
<xsd:element name="node" type = "nodeType"/>
<xsd:complexType name = "nodeType">
    <xsd:choice>
        <xsd:element name = "onOffReg" type="xsd:string"/>
        <xsd:element name = "selectionReg" type="xsd:string"/>
        <xsd:element name = "decisionReg" type="xsd:string"/>
        <xsd:element name = "divisionReg" type="xsd:string"/>
        <xsd:element name = "otherReg" type="xsd:string"/>
        <xsd:element name = "clockSource" type="xsd:string"/>
    </xsd:choice>
    <xsd:attribute name = "id" type = "xsd:ID" use = "required"/>
</xsd:complexType>
<!-- Edge and Edge Type -->
<xsd:element name="edge">
    <xsd:complexType>
        <xsd:simpleContent>
            <xsd:extension base = "edgeType">
                <xsd:attribute name = "id" type = "xsd:ID" use = "required"/>
                <xsd:attribute name = "from" type = "xsd:IDREF" use = "required"/>
                <xsd:attribute name = "to" type = "xsd:IDREF" use = "required"/>
            </xsd:extension>
        </xsd:simpleContent>
    </xsd:complexType>
</xsd:element>
<xsd:simpleType name = "edgeType">
    <xsd:restriction base = "xsd:string">
        <xsd:enumeration value = "MCLK"/>
        <xsd:enumeration value = "SMCLK"/>
        <xsd:enumeration value = "ACLK"/>
        <xsd:enumeration value = "CLK"/>
        <xsd:enumeration value = "TAOUT0"/>
        <xsd:enumeration value = "TAOUT1"/>
        <xsd:enumeration value = "TAOUT2"/>
        <xsd:enumeration value = "TBOUT0"/>
```

```xsd
            <xsd:enumeration value = "TBOUT1"/>
            <xsd:enumeration value = "CCIxA"/>
            <xsd:enumeration value = "CCIxB"/>
            <xsd:enumeration value = "CCI"/>
        </xsd:restriction>
    </xsd:simpleType>
</xsd:schema>
```

# D  Frequency Optimisation Graph Example (Subsection 3.3)

# E  Numerical Example for Jensen's Algorithm[4] (Subsection 4.1)

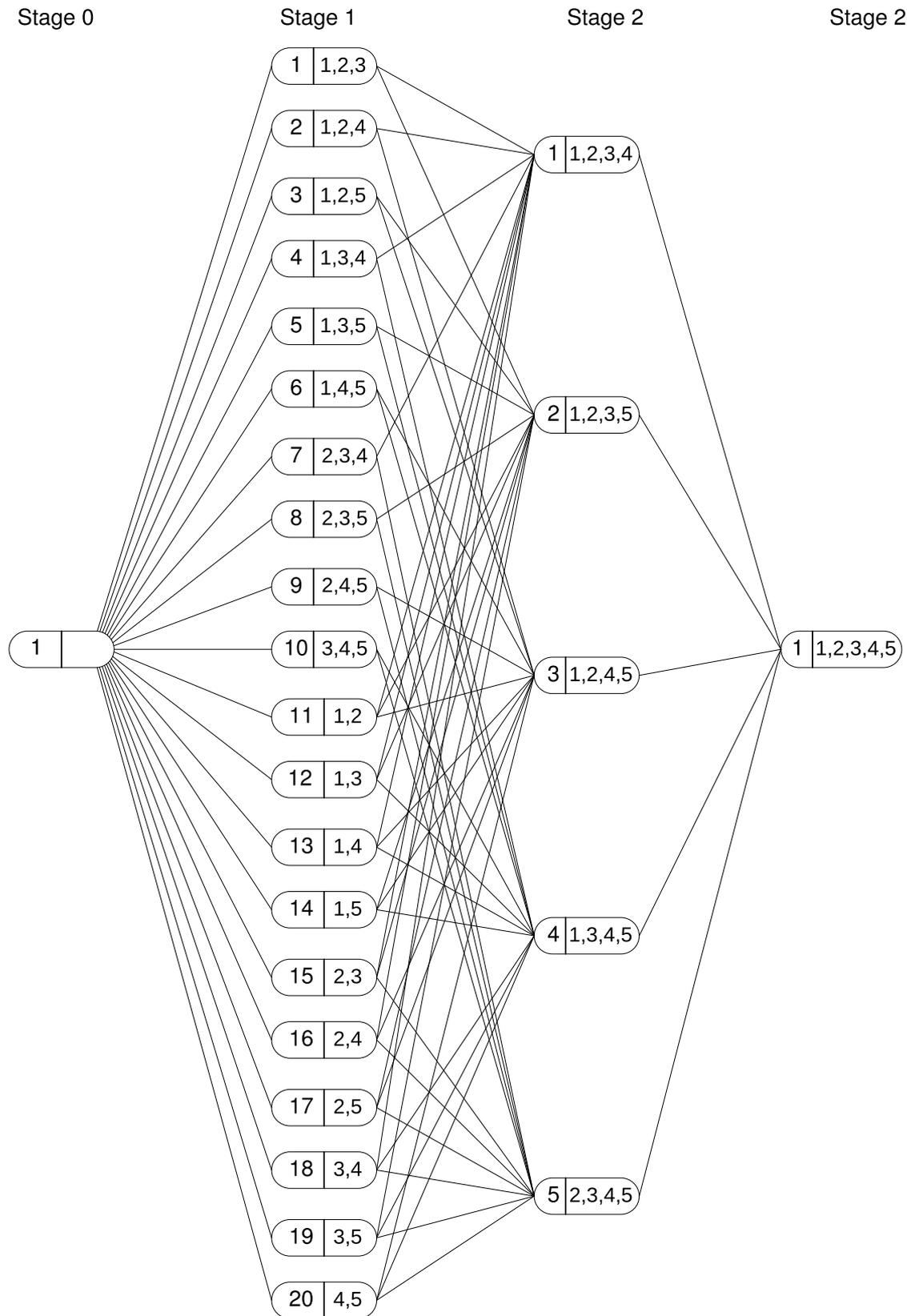

---
[4]The diagram is reproduced from [13]. The last *stage* is not considered as such by the author (hence the confusing label of "Stage 2"), since it forms single entity clusters, which do not add anything to the optimisation function.

# F Equations for TI MSP430's Basic Clock Module and Timer A (Subsection 4.2)

| | |
|---|---|
| $X$ | frequency of LFXT1CLK |
| $Y$ | frequency of XT2CLK |
| $Z$ | frequency of DCOCLK |
| $Y'$ | frequency of MCLK |
| $Z'$ | frequency of SMCLK |
| $X''$ | frequency of ACLK after divider |
| $Y''$ | frequency of MCLK after divider |
| $Z''$ | frequency of SMCLK after divider |
| $d_X$ | ACLK frequency divider |
| $d_Y$ | MCLK frequency divider |
| $d_Z$ | SMCLK frequency divider |
| $S, R$ | frequencies of Timer A external clock sources |
| $f_{TA}$ | frequency of Timer A clock before divider |
| $d_{TA}$ | Timer A frequency divider |
| $M_i$ | value of TACCRi register |
| $f_i$ | constants, frequencies of the timer generated by TACCRi |
| $a_i, b_i, c_i, d_{iX}, d_{iY}, d_{iZ}, d_{iTA}$ | binary parameters |

$$\begin{cases} Y'' d_Y = a_1 X'' d_X + a_2 Y + a_3 Z \\ Z'' d_Z = b_1 X'' d_X + b_2 Y + b_3 Z \\ 1 = a_1 + a_2 + a_3 \\ 1 = b_1 + b_2 + b_3 \\ d_X = d_{1X} \cdot 1 + d_{2X} \cdot 2 + d_{3X} \cdot 4 + d_{4X} \cdot 8 \\ d_Y = d_{1Y} \cdot 1 + d_{2Y} \cdot 2 + d_{3Y} \cdot 4 + d_{4Y} \cdot 8 \\ d_Z = d_{1Z} \cdot 1 + d_{2Z} \cdot 2 + d_{3Z} \cdot 4 + d_{4Z} \cdot 8 \\ 1 = d_{1X} + d_{2X} + d_{3X} + d_{4X} \\ 1 = d_{1Y} + d_{2Y} + d_{3Y} + d_{4Y} \\ 1 = d_{1Z} + d_{2Z} + d_{3Z} + d_{4Z} \\ f_{TA} = c_1 X'' + c_2 Y'' + c_3 S + c_4 R \\ 1 = c_1 + c_2 + c_3 + c_4 \\ f_{TA} = M_0 d_{TA} f_0 \\ f_{TA} = M_1 d_{TA} f_1 \\ f_{TA} = M_2 d_{TA} f_2 \\ d_{TA} = d_{1TA} \cdot 1 + d_{2TA} \cdot 2 + d_{3TA} \cdot 4 + d_{4TA} \cdot 8 \\ 1 = d_{1TA} + d_{2TA} + d_{3TA} + d_{4TA} \\ X \in \{0, 32768\} \\ Y \in \{0, 8000000\} \\ Z \in \{0, 750000, 1000000, 2000000, 4000000, 8000000\} \\ M_i \in [1, 65536] \end{cases}$$

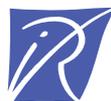